\documentclass[aps,prl,twocolumn,showpacs,superscriptaddress]{revtex4}

\usepackage{graphicx,amsmath,verbatim}
\pdfoutput=1
\newcommand{\be}{\begin{equation}}
\newcommand{\ee}{\end{equation}}

\begin{document}

\title{Resonant coupling of a Bose-Einstein condensate to a micromechanical oscillator}

\author{David Hunger}
\author{Stephan Camerer}
\author{Theodor W. H{\"a}nsch}
\affiliation{Fakult{\"a}t f{\"u}r Physik, Ludwig-Maximilians-Universit{\"a}t, Schellingstra{\ss}e~4, 80799~M{\"u}nchen, Germany}
\affiliation{Max-Planck-Institut f{\"u}r Quantenoptik,  Hans-Kopfermann-Str.~1, 85748~Garching, Germany}
\author{Daniel K{\"o}nig}
\author{J{\"o}rg P. Kotthaus}
\affiliation{Fakult{\"a}t f{\"u}r Physik, Ludwig-Maximilians-Universit{\"a}t, Schellingstra{\ss}e~4, 80799~M{\"u}nchen, Germany}
\author{Jakob Reichel}
\affiliation{Laboratoire Kastler Brossel, ENS/UPMC-Paris 6/CNRS, 24
  rue Lhomond, F-75005 Paris, France}
\author{Philipp Treutlein}
\email[To whom correspondence should be addressed. E-mail: ]{treutlein@lmu.de}
\affiliation{Fakult{\"a}t f{\"u}r Physik, Ludwig-Maximilians-Universit{\"a}t, Schellingstra{\ss}e~4, 80799~M{\"u}nchen, Germany}
\affiliation{Max-Planck-Institut f{\"u}r Quantenoptik,  Hans-Kopfermann-Str.~1, 85748~Garching, Germany}

\date{\today}

\begin{abstract}
We report experiments in which the vibrations of a micromechanical oscillator are coupled to the motion of Bose-condensed atoms in a trap. 
The interaction relies on surface forces experienced by the atoms at about one micrometer distance from the mechanical structure.
We observe resonant coupling to several well-resolved mechanical modes of the condensate. 
Coupling via surface forces does not require magnets, electrodes, or mirrors on the oscillator and could thus be employed to couple atoms to molecular-scale oscillators such as carbon nanotubes.
\end{abstract}


\maketitle

Ultracold atoms can be trapped and coherently manipulated close to a surface using chip-based magnetic microtraps (``atom chips'') \cite{Fortagh07}.
This opens the possibility of studying interactions between atoms and on-chip solid-state systems such as micro- and nanostructured mechanical oscillators \cite{Schwab05,Kippenberg08}.
Such resonators have attracted much attention e.g.\ due to the extreme force sensitivity \cite{Schwab05} down to the single spin level \cite{Rugar04} and the novel manipulation techniques demonstrated in cavity optomechanics \cite{Kippenberg08}.
The question is raised whether the sophisticated toolbox for coherent manipulation of the quantum state of atoms could be employed to read out, cool, and coherently manipulate the oscillators' state.
Several theoretical proposals have considered the coupling of micro- and nanomechanical oscillators to atoms \cite{Meiser06, Treutlein07, Genes08, Ian08, Hammerer09}, ions \cite{Wineland98,Tian04,Hensinger05}, and molecules \cite{Singh08}. 
They show that sufficiently strong and coherent coupling would enable studies of atom-oscillator entanglement, quantum state transfer, and quantum control of mechanical force sensors.
In most scenarios, the coupling relies on local field gradients, calling for very close approach of the atoms to the oscillator. 
In this respect, ground-state neutral atoms stand out because preparation \cite{Lin04} and coherent manipulation \cite{Treutlein04} at micrometer distance from a solid surface has already been demonstrated on atom chips. 
While the intrinsically weak coupling of neutral atoms to the environment enables long coherence times, it makes coupling to solid-state degrees of freedom non-trivial. So far, only first steps have been made to investigate coupling mechanisms experimentally.
Recently \cite{Wang06}, atoms in a vapor cell were magnetically coupled to a mechanical oscillator. There, thermal motion of the atoms limits the interaction time and the control over the coupling.

\begin{figure}[t]
\includegraphics[width=0.43\textwidth]{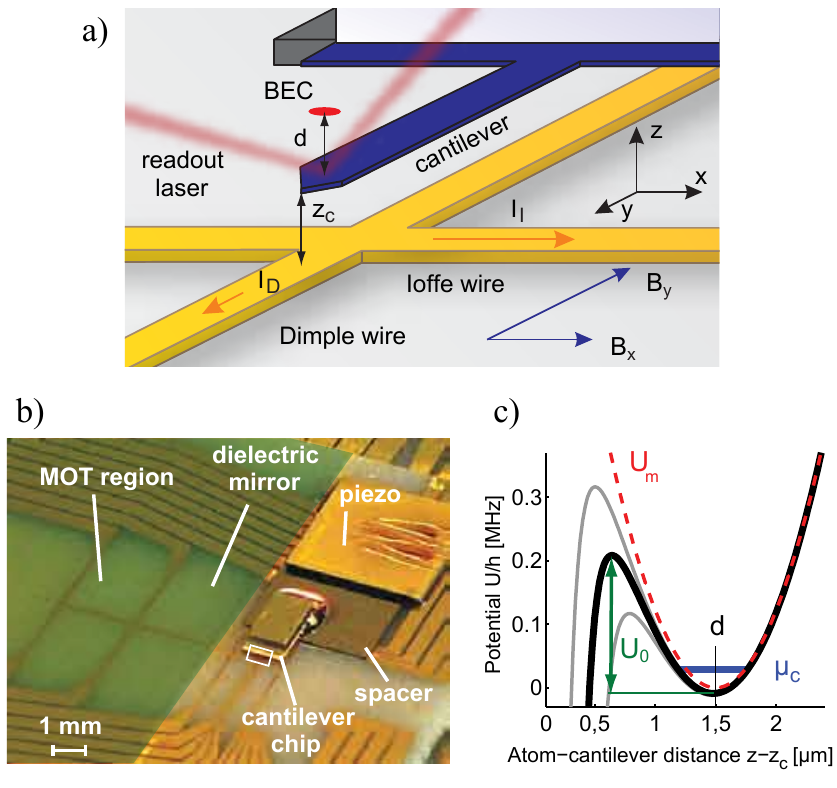}
\caption{\label{fig:Setup} (a) Micro-cantilever mounted on a chip with wires for magnetic trapping of atoms. Cantilever vibrations can be independently probed with a readout laser \cite{Supp}. (b) Photograph of the atom chip showing the magneto-optical trap (MOT) loading region and the cantilever subassembly with a piezo for cantilever excitation (scale bar: 1~mm). Rectangle: region shown in (a). (c) Potential $U=U_m+U_s$ for trap parameters as in Fig.~\ref{fig:DynLoss}a. Dashed red line: magnetic potential $U_m$. The surface potential $U_s$ reduces the trap depth to $U_0$. Cantilever oscillations modulate the potential, thereby coupling to atomic motion. Gray lines: $U$ during the extremum positions of the cantilever for an oscillation amplitude $a=120~$nm. Blue line: BEC chemical potential $\mu_c$ for $600$ atoms.
}
\end{figure}

In our experiment, we use a Bose-Einstein condensate (BEC) of $^{87}$Rb atoms \cite{Haensel01a} as a sensitive local probe for oscillations of a micromechanical cantilever.
Benefiting from its small spatial extent ($<300$~nm) and high positioning reproducibility ($<6$~nm) in a magnetic microtrap, we place the BEC at about one micrometer distance from the surface of the cantilever.
At such small distance, the magnetic trapping potential $U_m$ is substantially modified by the surface potential $U_s=U_\mathrm{CP}+U_\mathrm{ad}$. It consists of the Casimir-Polder (CP) potential $U_\mathrm{CP}$ \cite{Lin04,Harber05,Bender09} and an additional potential $U_\mathrm{ad}$ due to surface inhomogeneities or contamination \cite{Obrecht07b}. In the direction perpendicular to the surface, the combined potential is (see Fig.~\ref{fig:Setup}c)
\begin{eqnarray}
U[z] &=&  U_m + U_\mathrm{CP} + U_\mathrm{ad}  \nonumber \\
 &\approx& \frac{1}{2}m\omega_{z,0}^2(z-z_{t,0})^2  -\frac{C_4}{(z-z_c)^4} + U_\mathrm{ad}[z-z_c]. \nonumber
\end{eqnarray}
Here, $z_c$ is the position of the cantilever surface, $C_4$ the CP-coefficient, and $m$ the atomic mass. Like $U_\mathrm{CP}$, $U_\mathrm{ad}$ is attractive and quickly decays with atom-surface distance. 
The main effect of $U_s$ is to reduce the potential depth $U_0$ (see Fig.~\ref{fig:Setup}c) \cite{Lin04}. Additionally, it shifts the trap frequency from $\omega_{z,0}$ in the unperturbed magnetic trap to $\omega_z\approx(\omega_{z,0}^2+\frac{1}{m} \frac{\partial^2U_s}{\partial z^2})^{1/2}$ and the trap center from $z_{t,0}$ to $z_t\approx z_{t,0}-\frac{1}{m\omega_z^2}\frac{\partial U_s}{\partial z}$ \cite{Antezza04}.
When the cantilever oscillates, $U_s$ becomes time-dependent and leads to a modulation of $U_0$, $z_t$, and $\omega_z$ at the cantilever frequency $\omega_m$. We show that this excites atomic motion, which can be detected most simply via trap loss across the barrier $U_0$. The coupling depends strongly on the trap parameters and shows resonant behaviour if $\omega_m$ matches the frequency of a collective mechanical mode of the BEC. This can be used to control the interaction
efficiently.

Our setup integrates a SiN micro-cantilever of dimensions $(l,w,t)=(200,40,0.45)~\mu$m, Au/Cr metallized upper side, and fundamental resonance frequency $\omega_m/2\pi=10~$kHz on an atom chip (see Fig. \ref{fig:Setup}a,b and \cite{Supp}). Atoms can be trapped and positioned near the cantilever in a cigar-shaped, dimple-type Ioffe trap \cite{Haensel01a}.
We prepare BECs of typically $N = 2\times 10^3$ atoms in state $\left|F=2,m_F=2\right\rangle$ without discernible thermal component in a trap at a distance $d=16.6~\mu$m from the cantilever. At this distance we observe no influence of the surface.

\begin{figure}[t]
\includegraphics[width=0.47\textwidth]{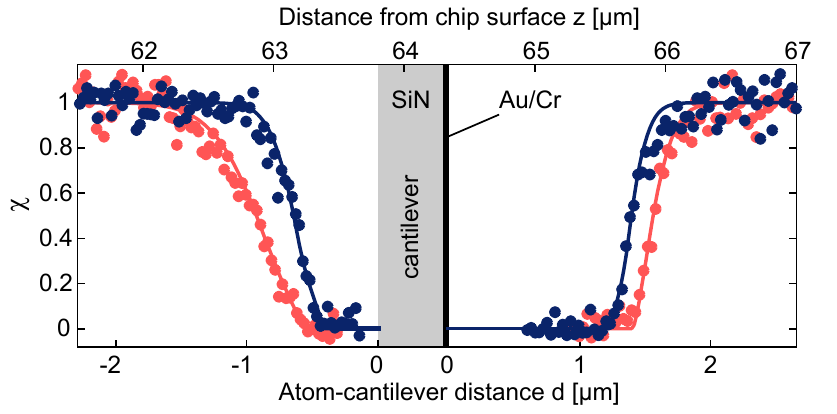}
\caption{\label{fig:SurfRamp} Fraction $\chi$ of atoms remaining in the trap after $t_h = 1$~ms at distance $d$ from the cantilever surface. Blue (red) datapoints correspond to a trap with $\omega_z/2\pi=10.0$~kHz ($5.1$~kHz). Solid lines: fit with a simple model \cite{Supp,Lin04}. The extracted cantilever position is shown. 
}
\end{figure}

As a prerequisite for dynamical coupling, we use a method similar to Ref.\ \cite{Lin04} to determine the range of atom-cantilever distances $d=z_{t,0} - z_c$ where the atoms are affected by $U_s$. In these measurements, the cantilever is undriven.
We compress the trapping potential to $\omega_z/2\pi=10~$kHz ($5~$kHz), resulting in a BEC radius of $290~$nm ($430~$nm), and ramp adiabatically within $1~$ms ($3~$ms) to a set value of $d$ close to the cantilever surface. The atoms are held there for an interaction time $t_h=1~$ms during which some of the atoms are lost because of the reduced $U_0$. The atoms are subsequently ramped back into a relaxed trap at large distance, where the remaining atom number $N_r$ is determined by absorption imaging \cite{Haensel01a}. 
Figure~\ref{fig:SurfRamp} shows the remaining fraction $\chi=N_r/N$ as a function of $d$.
The data shows that we can reproducibly prepare atoms at sub-micrometer distance from the cantilever.
We estimate the positioning reproducibility by measuring the atom number noise on the slope of a surface loss curve at $d=1.3~\mu$m. A worst case estimate that attributes all the noise to fluctuations of $z_{t,0}$ yields $\Delta z_{t,0}= 6~$nm r.m.s.

Taking advantage of the suspended structure, we can perform surface loss measurements on both sides of the cantilever, using the atoms as a ``caliper'' that measures the effective cantilever thickness including $U_s$.
Comparing the data with a simulation of $U$ allows us to calibrate $d$ to $\pm 160~$nm and to obtain information about $U_s$, because $U_m$ is very well known. 
In our analysis, we exploit that $\chi=0$ corresponds to the values of $d$ where the trap has vanished ($U_0 = 0$) in good approximation. 
Alternatively, we employ a model for the surface loss similar to \cite{Lin04} to describe the observed $\chi(d)$, yielding comparable results \cite{Supp}.
The data cannot be explained by $U_\mathrm{CP}$ alone but requires $U_\mathrm{ad} \gg U_\mathrm{CP}$ on at least one side of the cantilever. 
From measurements of dynamical atom-cantilever coupling (see below) we find that $U_s$ is significantly stronger on the metallized side. The data is consistently explained by potentials $U_\mathrm{ad} = -C_\mathrm{ad}/(z-z_c)^{4}$ with $C_\mathrm{ad}=(2\pm1)\times10^{2} \, C_4$ on the metallized side and $C_\mathrm{ad}=(10\pm10)\, C_{4,d}$ on the dielectric side. We use the coefficient $C_4=3\hbar c\alpha/(32\pi^2\epsilon_0)$ of a perfect conductor, with $\alpha=5.26 \times 10^{-39}~$Fm$^2$ the $^{87}$Rb ground state polarizability. On the dielectric side, $C_{4,d}=C_4\frac{\epsilon-1}{\epsilon+1}\Phi(\epsilon)$, with $\epsilon=4.0$ and $\Phi(\epsilon)=0.77$ for SiN \cite{Yan97}. A likely origin of $U_\mathrm{ad}$ is the stray field of surface adsorbates \cite{Supp,Obrecht07b}.

\begin{figure}[t]
\includegraphics[width=0.43\textwidth]{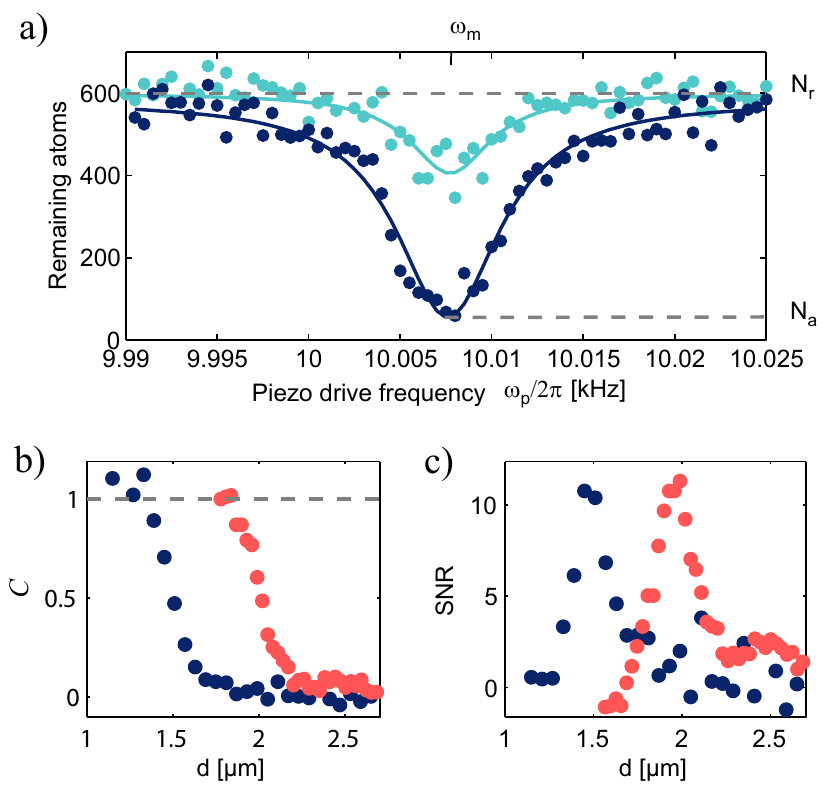}
\caption{\label{fig:DynLoss} (a) Remaining atom number after $t_h=3~$ms in a trap with $\omega_z/2\pi=10.5~$kHz at $d=1.5~\mu$m from the driven cantilever, for varying drive frequency $\omega_p$. The dark (light) blue circles correspond to a cantilever amplitude $a=120~$nm ($50~$nm) on resonance. Solid lines: Lorentzian fits with $6~$Hz FWHM, corresponding to the width of the cantilever resonance. (b+c) Contrast $C$ and signal to noise ratio SNR of the observed atomic signal as a function of $d$, for constant $a=90~$nm and $\omega_p = \omega_m$. Blue (red) datapoints correspond to $\omega_z/2\pi=10.5~$kHz ($5.0~$kHz) and $t_h=3~$ms ($20~$ms).
}
\end{figure}

We now describe our main experiments, where cantilever oscillations are coupled to the motion of the atoms nearby.
We excite the cantilever with the piezo at frequency $\omega_p$. When $\omega_p$ is resonant with the cantilever's fundamental out-of-plane mode at $\omega_m=2\pi \times10~$kHz, the cantilever oscillates with an amplitude $a$ of typically several tens of nm. We prepare BECs on the metallized side at $d=1.5~\mu$m in a trap with $\omega_z/2\pi = 10.5~$kHz, so that $\omega_z \approx\omega_m$, and let them interact with the vibrating cantilever. 
When $\omega_p$ is scanned from shot to shot of the experiment, a sharp resonance in the remaining atom number is observed for $\omega_p = \omega_m$, see Fig.~\ref{fig:DynLoss}a. 
This shows that we can use the atoms for cantilever readout.
Note that $a$ is more than one order of magnitude smaller than $d$, and the cantilever does not touch the atomic cloud. 
The surface potential of the oscillating cantilever modulates $z_t$ with an amplitude $\delta z_t=10$~nm ($4$~nm) for $a=120~$nm ($50~$nm), thereby exciting coherent motion of the atomic center of mass (c.o.m.). For large c.o.m.\ amplitudes the anharmonicity of the deformed trap and the reduced $U_0$ convert this motion into heating and loss.

Figure~\ref{fig:DynLoss}b shows the dependence of the atomic signal on $d$ for constant $a=90~$nm and $\omega_p = \omega_m$. 
We show the contrast $C=(N_r-N_a)/N_r$, where $N_a$ ($N_r$) is the remaining atom number with (without) resonant piezo excitation of the cantilever. We determine the signal visibility by the signal to noise ratio $\textrm{SNR}=(N_r-N_a)/\sigma$, with $\sigma= 32$ the overall noise observed without cantilever driving, see Fig.~\ref{fig:DynLoss}c.
The strong variation of the signal over a few hundred nm matches with the range of $d$ where $U_s$ modifies the trapping potential noticeably. We perform similar measurements in a trap resonant with a mode at $\omega_m=2\omega_z$ (see below) for longer $t_h$ and find comparable behaviour at larger distance. If we choose $d$ such that SNR is maximized, we observe a nearly linear dependence $C \propto a$ for $C<1$ and find $\delta z_t \propto a$ in the corresponding simulation. 
We observe the coupling on both sides of the cantilever.
Comparison of measurements at similar $U_0$ shows that $C/a$ is a factor of $3.2\pm0.6$ larger on the metallized side.
Because $C/a\propto \partial^2U_s/\partial^2 z$ this can be explained by a stronger $U_s$ on this side, and combined with loss measurements as in Fig.~\ref{fig:SurfRamp} we can quantitatively infer the strength of $U_s$ on both sides \cite{Supp}.

\begin{figure}[t]
\includegraphics[width=0.46\textwidth]{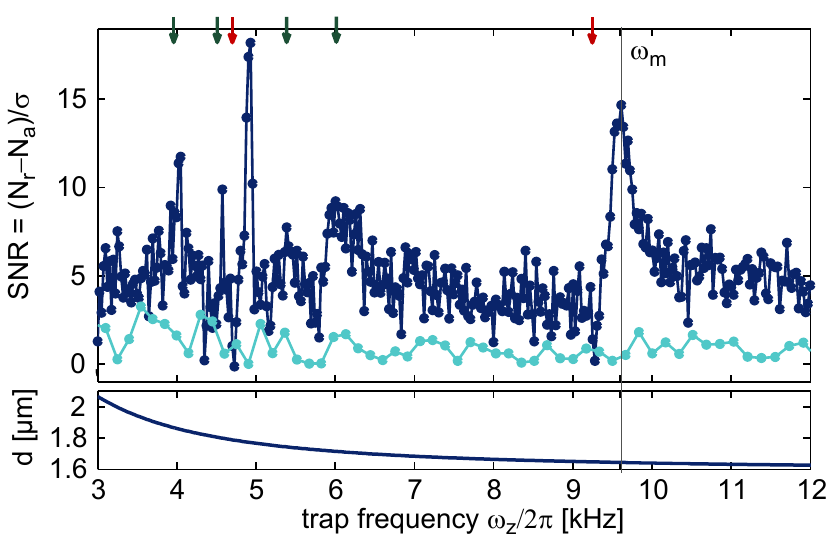}
\caption{\label{fig:Cantres} Top graph: BEC response as a function of $\omega_z$ for constant $a=180~$nm and $t_h=20~$ms (dark blue). Datapoints are connected to guide the eye. Light blue: reference measurement without piezo excitation. We observe two major resonances at $\omega_m=\omega_z$ and $\omega_m=2 \omega_z$, up to four smaller ones (green arrows), and reproducible anti-resonances (red arrows). Due to cantilever aging, $\omega_m/2\pi=9.68~$kHz in this measurement. Bottom graph: Set values of $d$, chosen such that $N_r(\omega_z) \approx \textrm{const.}$ ($N_r(10~\mathrm{kHz})=700$, $N_r(5~\mathrm{kHz})=1100$) and $N_a$ does not saturate.}
\end{figure}
The BEC can be regarded as a mechanical oscillator prepared in the quantum mechanical ground state.
Because of atomic collisions, it has a non-trivial spectrum of collective mechanical modes \cite{Stringari96,Kimura99}.
To demonstrate that the cantilever can be coupled selectively to different BEC modes, we measure the dependence of the atomic response on $\omega_z$.
The cantilever is excited to constant amplitude and coupled to the BEC on the metallized side. In Fig.~\ref{fig:Cantres} we show how the observed atomic SNR changes when we scan $\omega_z$.
The measured spectrum shows strong resonances at $\omega_m=\omega_z$ and $\omega_m=2\omega_z$. They correspond, respectively, to the atomic c.o.m.\ mode and the high frequency $m_l=0$ collective mode of the BEC in our cigar-shaped trap \cite{Stringari96,Kimura99}. In our trap, the latter coincides with the breathing mode of the thermal component of the gas.
The mode at $\omega_z$ ($2\omega_z$) is excited by the cantilever through modulation of $z_t$ ($\omega_z$) and we calculate a modulation amplitude of $\delta z_t=7~$nm ($\delta \omega_z=2\pi \times 150~$Hz). For the resonance at $\omega_m = 2\omega_z$, we observe a linewidth of only $60~$Hz, corresponding to a quality factor of $\approx 100$. 
Due to the trap anharmonicity, a thermal component can lead to a broadening of the resonances. This could explain the lineshape of the c.o.m.\ mode, where we expect a larger thermal component due to stronger heating at the higher trap frequency \cite{Supp}.
Next to the resonances, we observe reproducible ``anti-resonances'' where the atomic response is suppressed by a factor of $20$.
This can be used to switch the coupling on and off.
Yet, at this point, we have no clear explanation for their origin.
Furthermore, we find up to four weaker resonances at frequencies $\omega_m=(1.6,1.8,2.1,2.4)\,\omega_z$. The first resonance can be identified with the $|m_l|=2$ quadrupole mode of the BEC \cite{Stringari96}, whose frequency is given by  $\omega_m=\omega_z\sqrt{2(1+E_{\mathrm{kin},\perp}/E_{\mathrm{pot},\perp})}$ \cite{Kimura99}. We calculate the BEC kinetic energy $E_{\mathrm{kin},\perp}$ and potential energy $E_{\mathrm{pot},\perp}$ in the radial direction as in \cite{MunozMateo07} for $1100$ atoms, which reproduces the measured mode frequency. 
At smaller $d$, we observe broadening of the resonances, and the resonance at $\omega_m=1.6\,\omega_z$ becomes stronger than that at $\omega_m=2\,\omega_z$.

We have used trap loss as the simplest way to detect BEC dynamics induced by the coupling.
Measurements as in Fig.~\ref{fig:DynLoss}a yield a minimum resolvable r.m.s.\ cantilever amplitude of $a_\mathrm{rms}=13\pm4~$nm for SNR=1 without averaging.
This value is limited by the strong anharmonicity of the trap, and by the short trap lifetime of 18 ms (55 ms) for $\omega_z=2\pi \times 10$ kHz (5 kHz) due to three-body collisional loss and technical heating. 
Anharmonicity gives rise to dephasing and thereby limits the cloud amplitude for a given cantilever amplitude.
For trap loss to occur, the cantilever has to drive the BEC to large-amplitude oscillations with $\sim 10^3$ phonons. By contrast, BEC amplitudes down to the single phonon level could be observed by direct imaging of the motion.
A coherent state $| \alpha \rangle$ of the c.o.m.\ mode of $N=100$ atoms with $\alpha =1$ released from a relaxed detection trap with $\omega_z=2\pi\times 100$~Hz has an amplitude of $\sqrt{2\hbar\omega_z/mN} \alpha t=400$~nm after $t=4$~ms time-of-flight. This is about $10\%$ of the BEC radius and could be resolved by absorption imaging with improved spatial resolution.
Assuming that the coupling is linear in $a_\mathrm{rms}$, we estimate that $a_\mathrm{rms}=0.2$~nm would excite the BEC to $\alpha =1$ within $t_h=20$~ms and could thus be detected. This would allow to observe the thermal motion of our cantilever, which has a relatively large effective mass $M=5$~ng and correspondingly small r.m.s.\ thermal amplitude $a_\mathrm{th}=\sqrt{k_BT/M\omega_m^2}=0.4~$nm, where $T=300$~K is the cantilever temperature.

Our experiment is a first demonstration of mechanical coupling between a resonator and ultracold atoms.
The coupling relies on fundamental atom--solid state interactions and does not require fabrication of magnets, electrodes, or mirrors on the oscillator. It could thus serve as the connecting element between atoms and molecular-scale oscillators such as carbon nanotubes  \cite{Poncharal99,Babic03}.
A single-wall nanotube of $15~\mu$m length has $\omega_m/2\pi = 20$~kHz and $M=2\times 10^{-17}$~g, resulting in $a_\mathrm{th}=4~\mu$m at $T=300$~K and a quantum-mechanical ground state amplitude of $a_\mathrm{qm}=\sqrt{\hbar/2M\omega_m}=0.2~$nm. 
This could potentially be detected with the BEC. 
The surface potential of the nanotube is expected to be a factor $\sim 20$ weaker than $U_{\mathrm{CP}}$ of a bulk conductor \cite{Fermani07}, which could be compensated to some extent by closer approach of the atoms to the nanotube. 
Alternatively, electrostatic charging of the nanotube could increase the coupling. It is interesting to study whether such a coupled atom-nanotube system could approach the strong coupling regime \cite{Tian04}.
The degree of control over atoms close to a surface demonstrated here is an important ingredient also for coupling schemes that rely on functionalized cantilevers \cite{Treutlein07}.
Furthermore, extending the method of using atoms as a ``caliper'', ultracold atoms can serve as a three-dimensional scanning probe \cite{Aigner08}, permitting to map out weak electromagnetic fields and surface potentials even inside excavations.

We acknowledge helpful discussions with T. Steinmetz, I. Favero, P. B{\"o}hi, and M. Riedel.
This work was supported by the Nanosystems Initiative Munich. T.W.H.\ gratefully acknowledges support by the Max-Planck-Foundation.




\newpage

\onecolumngrid
\section{Resonant coupling of a Bose-Einstein condensate to a micromechanical oscillator: Auxiliary Material}
\twocolumngrid

\noindent This Auxiliary Material is organized as follows:

In Section I, we give more detailed information on our experimental setup.

In Section II, we discuss a model that describes for known surface potential $U_s$ how the atoms are lost when the trap is ramped towards the surface of the undriven cantilever (see solid lines in Fig.~2 of the paper). The model was originally developed in \cite{Lin04ap}. We discuss several refinements of the model compared to \cite{Lin04ap} and show that our analysis is independent of model details.

In Section III, we describe how we calibrate the atom-surface distance $d$ and obtain information about the surface potential $U_\mathrm{s}$. The analysis combines measurements of atom loss in $U_s$ as in Fig.~2 of the paper with measurements of the strength of the dynamical coupling as in Fig.~3.
We discuss Rb adsorbates on the cantilever surface as a likely explanation for the observed additional surface potential $U_\mathrm{ad}$. 
While this analysis provides further insight into the surface potential, we point out that the experiments on dynamical atom-cantilever coupling presented in our paper do not rely on a certain type of surface potential and do not require precise knowledge of $U_s$. The strong decay of $U_s$ with $d$, which is a general feature of surface potentials, allows one to adjust the strength of the interaction to a desired value by adjusting $d$.

In Section IV, we point out limitations due to collisional loss and technical heating observed in the attempt to approach the surface as closely as possible by using traps with higher trap frequency. We discuss how these limitations could be partially circumvented.

\section{I. Experimental setup}\label{sec:ExpSetup}
\subsection{Micro-Cantilever}
The oscillator in our experiment is a commercial SiN AFM cantilever of dimensions $(l,w,t)=(200,40,0.45)~\mu$m with a $65~$nm thick Au/Cr mirror on one side for optical readout.
Cantilever oscillations can be excited with a piezo. We calibrate the mechanical oscillation amplitude $a$ by implementing a laser beam deflection readout \cite{Putman92ap}. It involves a readout laser at $850~$nm that is focused onto and reflected from the cantilever tip. The angular deflection of the beam is detected with a quadrant photodiode and a lock-in amplifier. The readout has a sensitivity of $2\times10^{-12} ~ \textrm{m}/\sqrt{\textrm{Hz}}$, which allows us to resolve the thermal motion of the fundamental mode. Comparison with the driven cantilever amplitude yields a piezo driving efficiency of $80\pm15~$nm/Vpp on resonance.
We measure a cantilever frequency $\omega_m/2\pi=10~$kHz and a quality factor $Q=\omega_m/2\kappa = 3100$ for the fundamental out-of-plane mode, where $\kappa^{-1}$ is the $1/e$ amplitude decay time. A slow drift (over days) towards lower $\omega_m$ is observed. It does not depend on whether experiments are performed, and we attribute it to aging of the layered Au/Cr/SiN structure.
The cantilever chip is glued onto a spacer, ensuring a convenient separation of $z_c \approx 64~\mu$m between cantilever and atom chip surface. 

\subsection{Atom chip and BEC preparation}
The cantilever subassembly is glued onto an AlN chip with microfabricated gold wires for magnetic trapping of atoms (see Fig.~1 of the paper). The wires are defined photolithographically and grown to $5~\mu$m thickness in an electroplating process \cite{TreutleinThesis08ap}.
The chip forms the top wall of a glass cell vacuum chamber \cite{Du04ap} with $5\times 10^{-10}$~mbar background pressure. 
The atom chip has a loading region with a dielectric mirror to operate a mirror-magneto-optical trap (mirror MOT) \cite{Haensel01aap}, which collects $10^7$ atoms during $6~$s loading. After $3~$ms of optical molasses cooling the atoms are pumped into state $\left|F=2,m_F=2\right\rangle$ and trapped magnetically. The subassembly with the mechanical cantilever is located at a lateral distance of $6.5~$mm from the MOT center. To transport the cloud to the cantilever, we use a wire guide \cite{Fortagh07ap} with a superimposed axial quadrupole field from external coils, which can be shifted by a homogeneous field along the wire axis.

At the cantilever, the atoms are held in a cigar-shaped Ioffe trap, created by currents in a $50~\mu$m wide Ioffe wire and a crossing ``dimple'' wire in combination with homogeneous bias fields  \cite{Haensel01aap} (see Fig. 1 of the paper).  
The trap frequency $\omega_z$ can be widely adjusted (up to $\omega_z=2\pi\times20~$kHz) while maintaining $\omega_z \approx \omega_y \approx 10\, \omega_x$ for a large range of distances $z_{t,0}$ from the atom chip surface.
We perform radio-frequency evaporative cooling \cite{Ketterle99ap} with a first stage at $d=43~\mu$m from the cantilever surface. Pure BECs of typically $N=2.0\times 10^3$ atoms are then produced in a second stage at $d=16.6~\mu$m in a trap with $[\omega_x,\omega_z\approx\omega_y]=2\pi\times[0.6, 2.8]~$kHz, after an overall evaporation time of $1.9~$s. The lifetime in this trap is $2.5~$s, close to the pressure limited lifetime of $3.2~$s far away from surfaces.

Typical parameters for coupling to the cantilever are currents $(I_I, I_D)=(1.855, 0.400)~$A and homogeneous bias fields $(B_x, B_y)=(14.66, 59.21)~$G (see Fig. 1 of the paper), which results in $z_{t,0}=65.8~\mu$m and $\omega_{x,y,z}=2\pi\times(0.8,10.4,10.5)$~kHz. For this trap, $d=1.5~\mu$m, $U_0/h = 200$~kHz, and $z_{t,0}-z_t = 16$~nm. 
Three-body collisional loss and technical heating depend on trap frequency and limit the BEC lifetime to 18~ms (55~ms) for $\omega_z/2\pi=10$~kHz ($5$~kHz) close to the surface.

We calibrate $\omega_{z,0}$ in the different traps by trap modulation spectroscopy. We modulate the Ioffe current such that the trap minimum is modulated along the $z$-axis. For traps close to the cantilever this has a similar effect as the oscillating cantilever. We also use trap loss for readout in this measurement. The observed loss resonances have a relatively large width of $70-600~$Hz for $\omega_{z,0}/2\pi=3-14~$kHz, indicating trap anharmonicity. The measured trap frequency is thus not $\omega_{z,0}$, the value corresponding to the undisturbed trap, but corresponds to the inverse of the oscillation period for rather large amplitude oscillations in the deformed trap. As this is the relevant frequency for the coupling measurements, we refer to it as $\omega_z$. The relative uncertainty in the obtained $\omega_z$ is $<\pm 3~\%$, which is also the amount of anharmonicity (for a cloud oscillating up to the barrier a simulation yields $5\%$ anharmonicity). A simulation of the trapping potential (see chapter III) deviates from the measured frequencies by less than $^{+1}_{-5}\%$.

For measurements on the unmetallized backside of the cantilever, we transport thermal clouds around the cantilever and prepare the BEC there. The clouds are imaged after transporting back around the cantilever. We have no clear signature of condensation in this trap due to heating during transport. From a fit to the measurements in Fig. 2 in the paper (see below), we find that the temperatures are $<1.5\,T_c$, where $T_c$ is the critical temperature for BEC.

\section{II. Model for surface-induced atom loss}\label{sec:lossmodel}

Here we discuss the model for the loss of atoms in the attractive surface potential $U_s$ of the undriven cantilever (see solid lines in Fig.~2 of the paper). 
A simple model describing such measurements was developed in \cite{Lin04ap}. When the trapping potential is ramped to the surface, the trap depth $U_0$ is reduced by $U_s$. The model assumes that this leads to a sudden loss of the tail of the Boltzmann distribution of the residual thermal cloud at temperature $T$ coexisting with the condensate. Furthermore, it includes 1D evaporation from the trap at the reduced $U_0$ during the hold time $t_h$. According to this model, the remaining fraction of atoms in the trap is given by 
\begin{equation}
\chi=(1-e^{-\eta})e^{-\Gamma(\eta) t_h},
\end{equation}
where $\eta=U_0/k_B T$ is the ratio of the trap depth and the thermal energy and $\Gamma(\eta)=f(\eta)\exp(-\eta)/\tau_{el}$ is the evaporation rate. It contains the elastic collision time $\tau_{el}$ and a dimensionless factor $f(\eta)=2^{-5/2}(1-\eta^{-1}+\frac{3}{2}\eta^{-2})$, which accurately describes 1D evaporation for $\eta\geq 4$ \cite{Surkov96ap}. Evaporation is important when $t_h\gg\tau_{el}$. In the measurement of Fig.~2 of the paper, $\tau_{el}=0.2-0.6~$ms and $t_h=1~$ms, so that evaporation has only a small effect. Similarly, tunneling through the barrier, which can be accounted for by a small reduction of $U_0$, has only a minor effect for our parameters.
In comparing with the data, we leave $T$ as a free parameter. A fit to the data yields $T=(1.5,1.2,1.0,0.6)\, T_c$ (solid lines from left to right in Fig.~2), where $T_c$ is the critical temperature for BEC in the corresponding traps. These values are systematically larger but still in reasonable agreement with those from independent measurements of $T$ in time-of-flight for corresponding traps on the metallized side.

Although this simple model already describes our data fairly well, several improvements are possible:
 
(i) The model describes all atoms as being in the thermal cloud. A more accurate description can be obtained by assuming a bimodal cloud for $T<T_c$, with a thermal component of $N_\mathrm{th}=N(T/T_c)^3$ atoms and a condensate of $N_c = N-N_\mathrm{th}$ atoms \cite{Dalfovo99ap}. The condensate chemical potential is $\mu_c=\frac{\hbar\bar{\omega}}{2}(15N_c a_s/\bar a)^{2/5}$ and the critical temperature for BEC is given by $k_B T_c=0.94\,\hbar\bar{\omega}N^{1/3}$ \cite{Dalfovo99ap}. Here, $\bar{\omega}=(\omega_x \omega_y \omega_z)^{1/3}$ is the mean trap frequency, $a_s=5.4~$nm the scattering length, and $\bar a=\sqrt{\hbar/m\bar{\omega}}$ the mean oscillator length. The thermal cloud is lost for $U_0 \geq \mu_c$ as described above with a modified $\eta=(U_0-\mu_c)/k_BT$. The condensate is lost for $U_0<\mu_c$, where the number of remaining atoms $N_r$ can be determined from $\mu_c[N_r]=U_0$.
(ii) Most of the surface loss occurs for $\eta \leq 1$, where the simple evaporation law is no longer valid and leads to unphysically large $\Gamma$. We correct this by introducing a cutoff at the cross dimensional mixing rate \cite{Monroe93ap}, which we implement by setting $\Gamma^{-1}=\tau_{el}(\frac{1}{f(\eta)\exp(-\eta)}+2.7)$. 
(iii) The repulsive interaction between the condensate and the thermal cloud pushes the latter out of the trap center, which leads to a broadening of the loss curves.
This effect could be included by an effective potential $U_\mathrm{th}(\mathbf{r})=\left| \tfrac{1}{2}m(\omega_x^2 x^2 + \omega_y^2 y^2 + \omega_z^2 z^2)-\mu_c\right|$ for the thermal cloud.
(iv) Technical heating, three-body collisional loss, and cooling due to evaporation of the atoms are not included in the model. These effects have a strong dependence on trap frequency.

By implementing the above improvements (i) and (ii), we found that the resulting change in the calibration of the atom-cantilever distance $d$ is $\pm80~$nm, which is within our error bar on $d$ (see below).
Furthermore, we point out that the data can also be analyzed without a detailed model for atom loss by simply exploiting that $\chi=0$ corresponds to the values of the atom-cantilever distance $d$ where the trap has vanished (which is well described by the condition $U_0=0$). This analysis depends only on the knowledge of the trapping potential $U$, and again yields similar results as the model described above for the short $t_h$ of the measurements in Fig.~2, where evaporation does not play an important role.

\section{III. Distance calibration and analysis of the surface potential}\label{sec:distcal}

Here we describe in detail how we use the atoms to obtain information about the surface potential $U_s=U_\mathrm{CP}+U_\mathrm{ad}$ and a calibration of the atom-cantilever distance $d=z_{t,0} - z_c$ on both sides of the cantilever. The analysis combines information from measurements of atom loss in the surface potential as in Fig.~2 of the paper with measurements of the strength of the dynamical coupling as in Fig.~3.

A first rough estimate of the cantilever-chip distance $z_c=68\pm10~\mu$m is obtained from the knowledge of the spacer chip thickness ($48\pm5~\mu$m) on which the cantilever is mounted and the thickness of the glue layers. Here, $z_c$ refers to the position of the surface of the metallized side of the cantilever. The surface of the dielectric side is located at $z_c - t$, where $t=450\pm40~$nm is the cantilever thickness specified by the manufacturer and confirmed in electron microscope images.

For a more precise determination of $z_c$ we use several measurements of atom loss in the surface potential as shown in Fig.~2 of the paper.
In these measurements, the position of the magnetic trap minimum, $z_{t,0}$, is obtained from a simulation of the magnetic trapping potential $U_m$. Our simulation takes into account the finite width and length of the wires as well as the rectangular geometry of the three pairs of coils generating magnetic bias fields. We check the simulated $U_m$ by comparison with measurements of the trap frequencies, the magnetic field at the trap bottom, and the trap position in absorption images. From this we estimate a relative uncertainty in $z_{t,0}$ of $\pm 3\%$. This leads to an absolute uncertainty of $\pm 2~\mu$m at $z_{t,0} = 65~\mu$m. This is also the absolute uncertainty in the $z$-axis in Fig.~2 of the paper. 

A crucial point in our measurements is that we can approach the cantilever from both sides, using the atoms as a ``caliper''. Because the cantilever has to lie somewhere in the region where $\chi=0$ in Fig.~2, this allows us to determine the absolute cantilever position $z_c = 64.7 \pm 2.1~\mu$m with an uncertainty comparable to the uncertainty in $z_{t,0}$.
However, we point out that the uncertainty in $d$ is much smaller than the absolute uncertainties in $z_{t,0}$ and $z_c$. This is so because the distance between magnetic traps right above and below the cantilever is known to $\pm 60~$nm (corresponding to the $\pm 3\%$ relative uncertainty in $z_{t,0}$). For perfectly known surface potentials $U_s$, this would also be the uncertainty in $d$. 

The dominant contribution to the uncertainty in $d$ is due to the a priori unknown additional surface potential $U_\mathrm{ad}$. If we assume for the moment that only the CP-potentials are present on both sides of the cantilever (i.e.\ $U_\mathrm{ad}=0$ on both sides), we would expect from a simulation of $U=U_m + U_\mathrm{CP}$ that the ``effective cantilever thickness'' $t_\mathrm{eff}$, defined by the width of the window where $\chi=0$ in Fig.~2, is $t_\mathrm{eff}=1.4~\mu$m for $\omega_z/2\pi = 10$~kHz and $t_h=1~$ms. However, we observe $t_\mathrm{eff}=2.2~\mu$m. This shows that $U_s$ is significantly stronger than the expected contribution from $U_\mathrm{CP}$ on at least one side of the cantilever. We explain this by the presence of an additional potential $U_\mathrm{ad}$ due to surface inhomogeneities or contamination \cite{Speake03ap,Sandoghdar96ap,McGuirk04ap,Harber05ap,Obrecht07ap,Obrecht07bap}. Without taking into account further information about $U_\mathrm{ad}$, this leaves an uncertainty in $d$ of $\pm  400$~nm, corresponding to the difference between the observed and expected $t_\mathrm{eff}$.

The atoms could be used as a three-dimensional scanning probe that allows one to map out the spatial dependence of $U_\mathrm{ad}$ in detail and to determine whether it is due to magnetic, electrostatic, or other interactions, see e.g.\ the measurements in \cite{McGuirk04ap,Harber05ap,Obrecht07ap,Obrecht07bap}. As the characterization of $U_\mathrm{ad}$ is not the main focus of our present paper, we simply determine its strength relative to $U_\mathrm{CP}$ in the relevant range of $d$ by combining the measurements in Fig.~2 with information from measurements of dynamic atom-cantilever coupling as in Fig.~3 of the paper. 
We couple the atoms to the cantilever motion and measure the contrast $C$ of the atomic signal for several cantilever amplitudes $a$, using a trap with $\omega_z/2\pi=10~$kHz so that $\omega_m = \omega_z$. We find linear dependences both for $C \propto  a$ in the experiment (as long as $C<1$) and for $\delta z_t \propto a$ in the simulation of $U$, which implies a linear dependence $C\propto \delta z_t$.
Such measurements are performed on both sides of the cantilever in traps with similar $U_0$. 
We can determine $U_0$ to $10\%$ from the measured curves in Fig.~2 without detailed knowledge of $U_s$ or $d$.
Comparing measurements on both sides of the cantilever, we find that $C/a$ is a factor $\beta=3.2\pm0.6$ larger on the metallized side. Because of the observed linearity of the coupling, we conclude that $\delta z_t/a$ has to be larger by the same factor $\beta$. This implies a stronger surface potential on the metallized side. Stronger $U_s$ also implies larger $d$ to maintain the same $U_0$. Due to the fast decay of $U_s$ with $d$, a substantially larger $U_s$ is required on the metallized side (not just larger by a factor of order $\beta$).

A likely explanation for the observed $U_\mathrm{ad}$ are $^{87}$Rb adsorbates deposited during operation of the experiment. This effect was studied in detail in \cite{McGuirk04ap,Harber05ap,Obrecht07ap,Obrecht07bap}. The electric dipole moment of Rb on gold, $\mu_{el}\approx 1\times10^{-29}~$Cm, is about one order of magnitude stronger than on SiN \cite{McGuirk04ap,Obrecht07bap}. Furthermore, as most of the measurements are performed above the metallized surface, we estimate the adsorbed atom number on this side to be substantially larger than on the dielectric backside of the cantilever. Both effects would lead to a stronger $U_\mathrm{ad}$ on the metallized side.

The distance-dependence of the generated $U_\mathrm{ad}$ depends on the spatial distribution of adsorbates on the surface. The condensates from our cigar-shaped trap would result in an elongated distribution of adsorbates. This can be roughly approximated by a line of dipoles, for which the potential is given by 
\begin{equation}
   U_\mathrm{ad} = - \frac{\alpha}{2} \left| E \right|^2= -\frac{C_\mathrm{ad}}{(z-z_c)^{4}},
\end{equation}
where $\alpha$ is the $^{87}$Rb ground state polarizability and $E$ the electric field generated by the adsorbates \cite{McGuirk04ap}.

Our analysis is an iterative procedure in which we first choose a certain $C_\mathrm{ad}$ and then evaluate $\delta z_t$ and $d$ for the given $U_0$ on the dielectric side. This fixes the cantilever position $z_c$. Then we adjust $C_\mathrm{ad}$ on the metallized side to be consistent with the surface loss curves in Fig.~2 and extract $\delta z_t$ for the given $U_0$ on this side. We compare the values of $\delta z_t$ on both sides and start a new iteration with weaker (stronger) $C_\mathrm{ad}$ on the dielectric side if their ratio is smaller (larger) than the observed $\beta$, or finish if it equals the observed $\beta$.

The observed $\beta=3.2$ as well as the surface loss curves can be best explained by $C_\mathrm{ad}=(10\pm10)\, C_{4,d}$ on the dielectric side and $C_\mathrm{ad}=(2\pm1)\times10^{2} \, C_4$ on the metallized side.
The CP-coefficients $C_4$ and $C_{4,d}$ are given in the main text.
For these potentials, $z_c=64.36~\mu$m results.
The potential on the metallized side could be generated by $8\times10^6$ Rb atoms distributed over an area of $10\times 1~\mu$m$^2$, about two times the size of a condensate. This is a realistic atom number consistent with the number of experiments performed.

To check the robustness of our analysis against changes in the assumed distance-dependence of $U_\mathrm{ad}$, we performed similar analyses with other distance-dependences, such as $U_\mathrm{ad}\propto (z-z_c)^{-3}$ on both sides or $U_\mathrm{ad} \propto (z-z_c)^{-4}$ on the dielectric side and $U_\mathrm{ad}\propto (z-z_c)^{-3}$ on the metallized side. These analyses result in similar calibrations of $d$.
The overall error in $d$ is $\pm 160~$nm, which contains the uncertainty in $U_\mathrm{ad}$, $z_{t,0}$, $U_0$, $\beta$, as well as in the cantilever thickness, and a contribution due to residual oscillations of the atoms in the trap due to the ramping to the cantilever.

In our experiment, we observe that $U_\mathrm{ad}$ slowly changes over time by up to a factor of four on a time scale of weeks. This is consistent with the picture that atoms are deposited on the surface and subsequently diffuse or desorb again \cite{Obrecht07bap}. The measurements used to determine $U_\mathrm{ad}$ described above were all performed on the same day. The change in $U_\mathrm{ad}$ during the course of these measurements is negligible.

On the dielectric side, the thin SiN layer together with the Au/Cr film acts as a cavity or waveguide for the vacuum modes, which results in a correction to the CP-potential \cite{Curtois96ap}. At $d=1.0~\mu$m this leads to a $25$\% larger CP-potential, but only to a negligible shift ($<20~$nm) of $z_c$ or $d$.

\section{IV. Limitations on trap frequency and atom-surface distance}\label{sec:limitd}

In the attempt to approach the surface as closely as possible by using traps with higher $\omega_z$, we observe limitations due to three-body collisional loss and technical heating. 

Inelastic collisions lead to a density dependent loss rate $\gamma_{\mathrm{tbl}}=-L\left\langle n^2\right\rangle\propto\bar{\omega}^{12/5}N^{4/5}$ with $L=1.8\times10^{-41}$~m$^6/$s for $^{87}$Rb atoms in state $\left|F=2,m_F=2\right\rangle$ \cite{Soeding99ap}. For our coupling trap with $\omega_z/2\pi=10~$kHz ($5~$kHz) and a BEC atom number $N=2\times 10^3$, the mean density is $\langle n \rangle = 2.4\times10^{15}$~cm$^{-3}$ ($1.4\times10^{15}$~cm$^{-3}$), giving rise to $\gamma_{\mathrm{tbl}}=120~$Hz ($40~$Hz).
Three-body loss restricts experiments to low atom number and detains from higher trap frequencies. One way of circumventing this to some degree is to use pancake-shaped traps, e.g.\ an optical trap, which has reduced density at a given $\omega_z$.
Furthermore, we point out that by using single atoms instead of BECs for atom-cantilever coupling, collisional loss can be circumvented. With single atoms, trap frequencies of the order of $1~$MHz are possible \cite{Reichel02ap}, which would allow for $d \sim 100$~nm near a thick metallic object. Near nanoscale objects with weaker surface potentials, even smaller $d$ could be achieved.

Technical current noise leads to fluctuations of the trap position $z_t$ and the trap frequency $\omega_z$, resulting in heating \cite{Gehm98ap}. The fluctuations in $z_t$ result in a linear increase in atom cloud temperature that scales with $\omega_z^4$, while fluctuations in $\omega_z$ lead to an exponential temperature increase that scales with $\omega_z^2$. For a trap far from the surface with $\omega_z/2\pi=10~$kHz ($5~$kHz) we observe thermalization of condensates within $2~$ms ($6~$ms). At an atom-surface distance of $d=1.6~\mu$m ($d=2~\mu$m), where $U_0$ is decreased to $h \times 270~\mathrm{kHz}=7.5\,\mu_c$ ($h \times 100~\mathrm{kHz}=5.1\,\mu_c$), the measured condensate lifetime increases and is no longer limited by thermalization. We attribute this to the cooling effect of surface evaporation. Here, one is rather limited by atom loss, and we observe a trap lifetime of $18~$ms ($55~$ms).
The loss and heating rates impose a limit on the hold time $t_h$ close to the surface. We observe an optimum in $t_h$ that depends on the trap geometry and current noise level. Optimization of the trap geometry and reduction of current noise would enable smaller $d$ and larger $t_h$, thus increasing the atom-cantilever coupling strength and the observed signal. 

We find no influence of atom loss due to thermal magnetic near-field noise in our experiment. Such noise is generated by the Johnson current noise in conducting materials and can drive atomic spin flips, leading to loss of atoms from the trap \cite{Fortagh07ap}. Due to the small amount of metallic material on the cantilever, this mechanism does not reduce the trap lifetime to below $1~$s at $d=1~\mu$m and is negligible compared to the observed pressure-limited background loss rate for $d>1.4~\mu$m.


\end{document}